\documentclass[12pt,preprint]{aastex}

\def\rmit#1{{\it #1}}              

\def\etc{\rmit{etc.}}
\def\ie{\rmit{i.e.}}
\def\eg{\rmit{e.g.}}

\def\kms{\hbox{km$\;$s$^{-1}$}}

\shorttitle{MHS sunspot model} \shortauthors{Khomenko and
Collados}

\begin{document}


\title{MHS sunspot model from deep sub-photospheric to chromospheric layers}

\author{E. Khomenko\altaffilmark{1,2}, M. Collados\altaffilmark{1}}
\email{khomenko@iac.es, mcv@iac.es}

\altaffiltext{1}{Instituto de Astrof\'{\i}sica de Canarias, 38205
La Laguna, Tenerife, Spain} \altaffiltext{2}{Main Astronomical
Observatory, NAS, 03680, Kyiv, Ukraine}

\begin{abstract}
In order to understand the influence of magnetic fields on the
propagation properties of waves, as derived from different local
helioseismology techniques, forward modeling of waves is required.
Such calculations need a model in magnetohydrostatic equilibrium
as initial atmosphere to propagate oscillations through it. We
provide a method to construct such a model in equilibrium for a
wide range of parameters to be used for simulations of artificial
helioseismologic data. The method combine the advantages of
self-similar solutions and current-distributed models. A set of
models is developed by numerical integration of magnetohydrostatic
equations from the sub-photospheric to chromospheric layers.
\end{abstract}


\section{Introduction}

In the recent years, local helioseismology has provided new
insights into the sub - photospheric structure of quiet and active
regions of the Sun \citep{Duvall+etal1993, Kosovichev1999,
Kosovichev2002, Kosovichev+etal2000, Zhao+Kosovichev2003,
Braun+Lindsey2000}.
However, the influence of magnetic fields on data interpretation
has not been fully explored. Theoretical efforts have been made by
\citet{Crouch+Cally2003}, \citet{Cally2005, Cally2006},
\citet{Schunker+Cally2006}, \citet{Cally+Goossens2007},
\citet{Schunker+etal2008} in order to include mode conversion and
to model ray path of waves in magnetized structures by means of
analytical theory.
These studies confirm the potential importance on helioseismic
measurements of the so called surface effects caused by the
presence of a magnetic field. A more complete understanding of the
problem will, probably, be better reached via direct forward
modeling of helioseismological data, since magnetic fields of
arbitrary configuration may be used.
There are several recent works reported in this direction as, \eg,
\citet{Gizon+etal2006, Khomenko+Collados2006,
Parchevsky+Kosovichev2007, Shelyag+etal2007, Hanasoge2008,
Cameron+etal2008}. In all the works cited above
\citep[except][]{Shelyag+etal2007}, the authors apply a similar
strategy. In particular, they assume the existence of an
equilibrium atmosphere containing a magneto-static structure whose
properties may resemble to a larger or lesser extent those of a
sunspot or a magnetic flux tube. A small-amplitude perturbation is
then applied to the system in order to study wave propagation,
wave mode transformations, amplitude and phase behaviour of waves
in a complex magnetic field topology, \etc

The behaviour of waves observed in active regions is very
sensitive to their magnetic field configuration. Photosphere and
low chromosphere are regions where a small change of parameters
such as the size of the magnetic structure, or its temperature,
density or magnetic field strength and inclination, may produce
significant changes in the resulted wave field. Simulations can be
of invaluable help to explore, within a full parameter space,
different magneto-static structures in order to understand the
effects produced by the magnetic field on the measurable variables
used in local helioseismology.

The latter task requires a robust procedure to construct
magneto-static structures of desired properties. In this paper, we
propose a strategy with that aim  and apply it to obtain thick
structures, as prototypes for solar spots and pores.
As a minimum requirement, the model should fulfill the following
properties: (i) in the photosphere the model should, on average,
reproduce the properties of a typical sunspot; (ii) at the border,
the model should smoothly merge into a quiet-Sun non-magnetic
model atmosphere; (iii) there should be a possibility to choose
the profile of thermodynamic parameters at the sunspot axis;
Wilson depression should be taken into account; (iv) magnetic
field strength, inclination and the radius of the structure should
be adjustable; (v) the model should be easily extensible into an
arbitrary depth below the photosphere.

There is a vast amount of works on magneto-static models reported
in the literature. Leaving apart small-scale flux tube models,
those for thick structures can be divided into those possessing a
current sheet \citep[\eg][]{Pizzo1990}, with a sharp magnetic-non
magnetic interface, and those with distributed currents
\citep[\eg][]{Pizzo1986}, showing a smooth transition. Without
discussing advantages and disadvantages of the both, we will
proceed here with current-distributed models.

Present current-distributed models apply two different
philosophies. In the first category of models, the magnetic
structure is prescribed and the distribution of thermodynamic
variables is looked for to be in agreement with this structure.
This is the class of self-similar models, proposed by
\citet{Schluter+Temesvary1958} and then extended by, \eg,
\citet{Low1975, Low1980}. In the second category, the pressure
distribution is prescribed as the boundary condition at the axis
of the magnetic structure and in the far-away non-magnetic
atmosphere. Both, pressure and magnetic field, are iteratively
changed in the remaining points to reach an equilibrium situation
\citep{Pizzo1986}.

From the point of view of the requirements set above, both classes
of models have advantages and disadvantages. The approach by
\citet{Pizzo1986} is more fruitful in the photosphere, since the
pressure distributions of the field-free and magnetized
atmospheres can be taken from observations and are relatively well
known. However, for deep sub-photospheric layers, the models that
can be taken as boundary conditions are scarce. More precisely,
the quiet-Sun non-magnetic pressure stratification can be taken
from helioseismological data, for example, from the standard solar
model of \citet{Christensen-Dalsgaard+etal1996}. As for the
sunspot axis, no precise data are available \citep[see,
however][]{Zhao+Kosovichev+Duvall2001, Kosovichev2002,
Couvidat+Birch+Kosovichev2006}. The model of \citet{Pizzo1986}
turns out to be very sensitive to the pressure deficit inside
sunspots and the method, in general, has poor convergence if the
simulation box is too deep and is very sensitive to the guess of
the pressure distribution at the sunspot axis. This makes the
method unsuitable for the purpose of our work.

On the other side, the procedure proposed by \citet{Low1980} works
better in deep layers, where the gas pressure dominates over the
magnetic pressure. In the photosphere, where the plasma becomes
magnetically-dominated, negative pressures are frequently obtained
from the method of \citet{Low1980}. It is complicated to guess the
parameters of the magnetic field configuration in order to avoid
this problem. At the same time, if one wishes to extend the models
into the photosphere and higher layers, the magnetic field
strength is limited to rather low flux tube-like values, not
appropriate for sunspots \citep[see][]{Hanasoge2008,
Cameron+etal2008}.

In this paper, we take advantage of both Pizzo-like and Low-like
approaches and propose a method to calculate the magneto-static
equilibrium of a thick sunspot-like structure with the properties
defined above. Below we describe the equations that allow to
successfully merge results from both methods and show examples of
MHS solutions for a wide range of parameters. The conclusions are
given in the last section.

\section{Method}

We solve the equilibrium force balance equation together with
divergence-free condition for the magnetic field:
\begin{eqnarray}
 \label{eq:initial}
 - & \vec{\nabla}P & + \rho\vec{g} +
\frac{1}{4\pi}(\vec{\nabla}\times\vec{B})\times\vec{B} = 0 \,, \\
\nonumber
 & \vec{\nabla}\vec{B}  & = 0 \,.
\end{eqnarray}
Following \citet{Pizzo1986}, the equations are solved in
cylindrical coordinates $(r,\phi,z)$ and axial symmetry is assumed
(\ie, all variables are independent of $\phi$).
Under these conditions the magnetic field vector can be
conveniently written in terms of the field line constant $u$:
\begin{equation}
 \label{eq:bvec}
\vec{B}=\left( - \frac{1}{r}\frac{\partial{u}}{\partial{z}},
\frac{G(u)}{r}, \frac{1}{r}\frac{\partial{u}}{\partial{r}} \right)
\,,
\end{equation}
where $G(u)$ is a function related to the twist component of the
field.
The variable $u$ is used both in \citet{Pizzo1986} and
\citet{Low1980}.
The difference is that, in \citet{Low1980}, the analytical
expression for $u$ is postulated, while, in \citet{Pizzo1986}, the
shape of the field lines $u$ is looked for by solving iteratively
the equation of the force balance for given boundary conditions in
agreement with some pressure distribution.
Except for constants, the functional form of $u$ used by
\citet{Pizzo1986} as initial condition at the lower boundary of
the computational domain is exactly the same as the one postulated
by \citet{Low1980}.
Thus, both models can be joined in a natural way, assuming that
the deep layers of the model sunspot can be approximated by the
self-similar solution and the upper layers by the solution of
Pizzo.
%


In \citet{Low1980}, following the spirit of self-similar
solutions, the field line constant $u$ is expressed as a function
of one variable, $\varphi$:
\begin{equation}
u(r,z)  =  u(\varphi); \,\,\,  \varphi = r^2 \cdot F(z); \,\,\,
F(z) = (z^2 + a^2)^{-1} \,,
\end{equation}
where $a$ is a constant parameter.
The field is untwisted and the azimuthal component, $B_{\phi}$, is
zero. This is equivalent to setting $G(u)=0$  in
Eq.~\ref{eq:bvec}. Using the above expression, Eq.~\ref{eq:bvec}
can be rewritten as a function of $\varphi$:
\begin{equation}
\label{eq:low} \vec{B}=\left( - r \frac{d F(z)}{d z} \frac{d u}{d
\varphi} ,\, 0,\, 2F(z)\frac{d u}{d \varphi} \right) \,.
\end{equation}
Following Low, the function $d u/d \varphi$ has to satisfy certain
normalizations in order to fulfill the force balance equation.
This leads to the following expression:
\begin{equation}
\label{eq:u} \frac{d u}{d \varphi} =B_0^L h^2 \cdot \exp(-\eta
\varphi) \,.
\end{equation}
Here, $B_0^L$ is a parameter that controls the magnetic field
strength and $h$ is a suitable length scale (note that, in the
original paper of Low, dimensionless variables are used, while
here we choose to use physical dimensions for all the variables).

Introducing Eq.~\ref{eq:u} into Eq.~\ref{eq:low}, the horizontal
and vertical components of the magnetic field vector in the Low's
model are written as:
\begin{equation}
\label{eq:brlow} B_r (r,z)=2 B_0^L  \frac{(z-z_{\rm d})rh^2
}{((z-z_{\rm d})^2 + a^2)^2}  \exp \left( \frac{-\eta r^2}
{(z-z_{\rm d})^2 + a^2} \right)
\end{equation}
\begin{equation}
\label{eq:bzlow} B_z (r,z) =2 B_0^L \frac{h^2}{(z-z_{\rm d})^2 +
a^2} \exp \left( \frac{-\eta r^2} {(z-z_{\rm d})^2+ a^2} \right)
\end{equation}
The parameter $z_{\rm d}$ is a reference height, where the
magnetic field is purely vertical. The latter expression is
directly comparable with the one used in Pizzo as a boundary
condition at the bottom boundary of the domain:
\begin{equation}
\label{eq:bzpizzo} B_z(r,z_0) = B_0^P \exp
\left(-\frac{r^2}{r_e^2} \right)
\end{equation}
Comparing these two expressions, we see that,  if both models are
to be joined at some arbitrary height, $z=z_0$, the parameters of
the models should be related as:
\begin{equation}
\label{eq:ltp1} B_0^P = B_0^L \frac{2h^2}{(z_0-z_{\rm d})^2 + a^2}
\end{equation}
\begin{equation}
\label{eq:ltp2} r_e^2=((z_0-z_{\rm d})^2 + a^2)/\eta
\end{equation}

Keeping this in mind, the model can be constructed following the
steps described below.

\subsection{Step 1: Generation of a self-similar solution in deep layers}

In deep sub-photospheric layers, we calculate a self-similar
solution for $\vec{B}$ after Eqs.~\ref{eq:brlow} and
\ref{eq:bzlow}. The pressure and density distributions with height
and radius are found from analytical expressions \citep[equations
50 and 51 in][]{Low1980}. As a boundary condition at the right
boundary (field-free atmosphere) we take the pressure and density
from the model S of \citet{Christensen-Dalsgaard+etal1996}. The
choice of the field-free pressure and density is rather arbitrary,
however.
Because of azimuthal symmetry, the left boundary of the domain
corresponds to $r=0$, i.e., the axis of the magnetic structure.
The lower boundary is taken exactly at height $z_{\rm d}$, where
$B_r(r,z_{\rm d}) = 0$ at all distances $r$. In all the cases
presented here, $z_{\rm d} = -10$ Mm and the origin for the z-axis
is taken at the base of the photosphere.
Given pressure and density, the temperature distribution in the
model sunspot can be calculated using the equation of state either
in tabular form or the one for an ideal gas.
The parameters $\eta$, $a$ and $B_0^L$ can be chosen freely. An
additional free parameter is the height, $z_0$, that limits the
upper boundary of the self-similar model. Depending on the height
of the upper boundary, $B_0^L$ can be larger or smaller in order
to prevent from getting negative gas pressures.

\begin{figure}
\centering
\includegraphics[width=9cm]{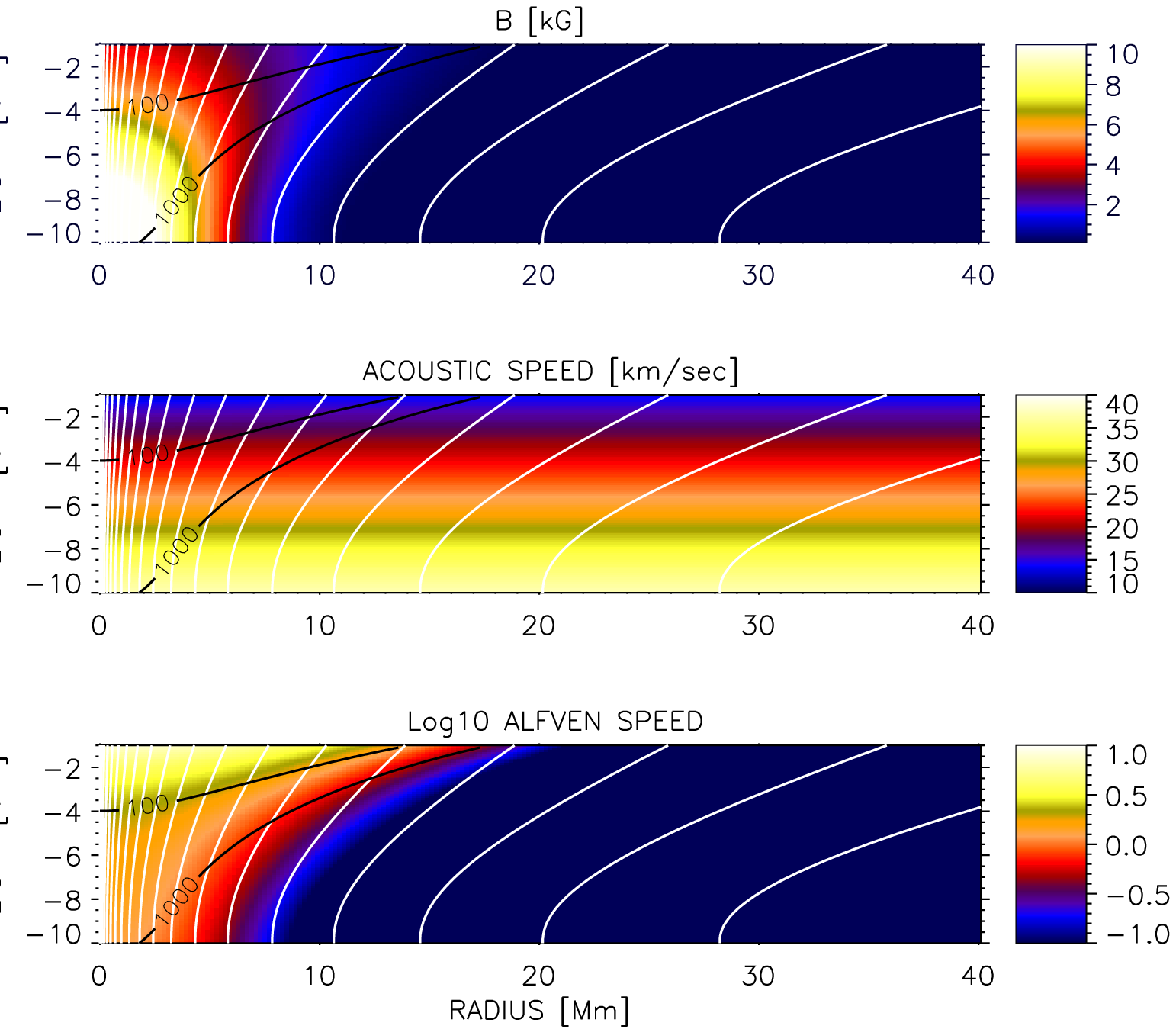}
\caption{Topology of the Low solution with $a= 2h$, $h=3$ Mm,
$\eta=1.3$, $B_0=25000$ G and $z_0=-1$ Mm. Top: magnetic field
strength; middle: acoustic speed; bottom: log of the Alfv\'en
speed. White lines are magnetic field lines.  Black lines with
labels are the contours of the ratio of the sound speed and the
Alfv\'en speed squared, $c_S^2/v_A^2$.} \label{fig:low}
\end{figure}

The basic topology of the solution is given in Fig.~\ref{fig:low},
for the parameters indicated in the figure caption. These
parameters are chosen on purpose in order to demonstrate that the
method is able to deal with large field strengths.
The sunspot radius at the bottom boundary is roughly defined by $a
\eta^{-1/2}$ \citep{Low1980}. The inclination of the field at the
top boundary changes with the distance from the axis, from 0 to
about 70 degrees at the right-most point of the domain. The
magnetic field is concentrated inside the first 15 Mm from the
axis, being weak in the rest of the domain. The field strength
drops at the axis from 12 kG at $z=-10$ Mm to 4 kG at $z=-$1 Mm
depth.
The gas pressure is always above the magnetic
pressure.
%
%
The Wilson depression is rather weak and the model sunspot is
almost thermally plane-parallel in deep layers, as follows from
the distribution of the acoustic speed, $c_S$. We do not have
criteria to judge how realistic is this description.  Data of the
sub-photospheric distribution of the sound speed in sunspots are
scarce and uncertain \citep[see, however, time-distance analysis
results by][]{Zhao+Kosovichev+Duvall2001, Kosovichev2002,
Couvidat+Birch+Kosovichev2006}.

\subsection{Step 2: Generation of potential solution in the overlaying atmosphere}

Given the values of $\eta$, $a$, $B_0^L$, $z_{\rm d}$, and $z_0$
we calculate the initial parameters of the Pizzo model from
Eqs.~\ref{eq:ltp1} and \ref{eq:ltp2}. This gives us $B_0^P=4$ kG
and $r_e=9.4$ Mm.
We follow the same steps as in the original paper of
\citet{Pizzo1986} and start from computing the potential solution:
\begin{equation}
\label{eq:potencial} \frac{\partial^2 u}{\partial r^2} -
\frac{1}{r}\frac{\partial u}{\partial r} + \frac{\partial^2
u}{\partial z^2} =0
\end{equation}
The bottom boundary of the domain coincides with the top boundary
from the previous step and is located at $z=-1$ Mm, below the
photosphere.
The field line constant $u$ at the bottom boundary is approximated
by:
\begin{equation}
u=r_e^2 B_0^P \left(1-\exp \left(-\frac{r^2}{r_e^2} \right)
\right)/2
\end{equation}
At the left (sunspot axis) and top boundaries $u=0$ (vertical
field) and $u$ approaches a constant value at the right boundary
(horizontal field).
With this set of boundary conditions, the boundary value problem
posed by Eq.~\ref{eq:potencial} is solved by standard methods.


\subsection{Step 3: Generation of magneto-static solution in the overlaying atmosphere}

The potential solution obtained in Step 2 is used as initial guess
in the integration of the complete force balance equation along
the magnetic field lines (equation 4 in the paper by Pizzo):
\begin{equation}
\label{eq:pizzo} \frac{\partial^2 u}{\partial r^2} -
\frac{1}{r}\frac{\partial u}{\partial r} + \frac{\partial^2
u}{\partial z^2} = -4 \pi r^2 \frac{\partial P(u,z)}{\partial u}
\end{equation}

In order to start iterations we need an approximation for the
distribution of pressure along the magnetic field lines $P(u,z)$.
Following \citet{Pizzo1986} and \citet{Low1975} we take it of the
form:
\begin{equation}
P(u,z)=P_0(u)\exp\left(-\int^z_0 \frac{dz^{'}}{h(u,z^{'})} \right)
\,,
\end{equation}
where $P_0(u)$ is the gas pressure along the bottom boundary. The
function $h(u,z)$ is a scale height.
For a complete description of the problem, the representative
pressure distributions along the axis and in the field-free quiet
atmosphere need to be specified.
As field-free atmosphere, we use of the model S of
\citet{Christensen-Dalsgaard+etal1996}, smoothly joint to the
VAL-C model of the solar chromosphere
\citep{Vernazza+Avrett+Loeser1981}.
At the axis, we use the \citet{Avrett1981} model in the upper
layers. In deep layers, we take a model by
\citet{Kosovichev+etal2000}, obtained from helioseismic inversions
of the sound speed beneath sunspots. This model already takes into
account the Wilson depression, which is about 450 km. The complete
model on the axis can be shifted up or down, though, if smaller or
larger values of the Wilson depression are required.
The model has a cool region just below the surface and a hot one
below, down to about -10 Mm. For our purposes we take this model
starting from -1 Mm depth and, thus, the hot layer is not taken
into account.
Once these models are specified, we calculate a smooth transition
between them for the gas pressure $P(u,z)$ and scale height
$h(u,z)$ distributions, as given by \citet{Pizzo1986} in his
equations (13), (17) and (18). Then, the Eq.~\ref{eq:pizzo} is
iterated until a convergence criterion is reached.

\begin{figure}
\centering
\includegraphics[width=9cm]{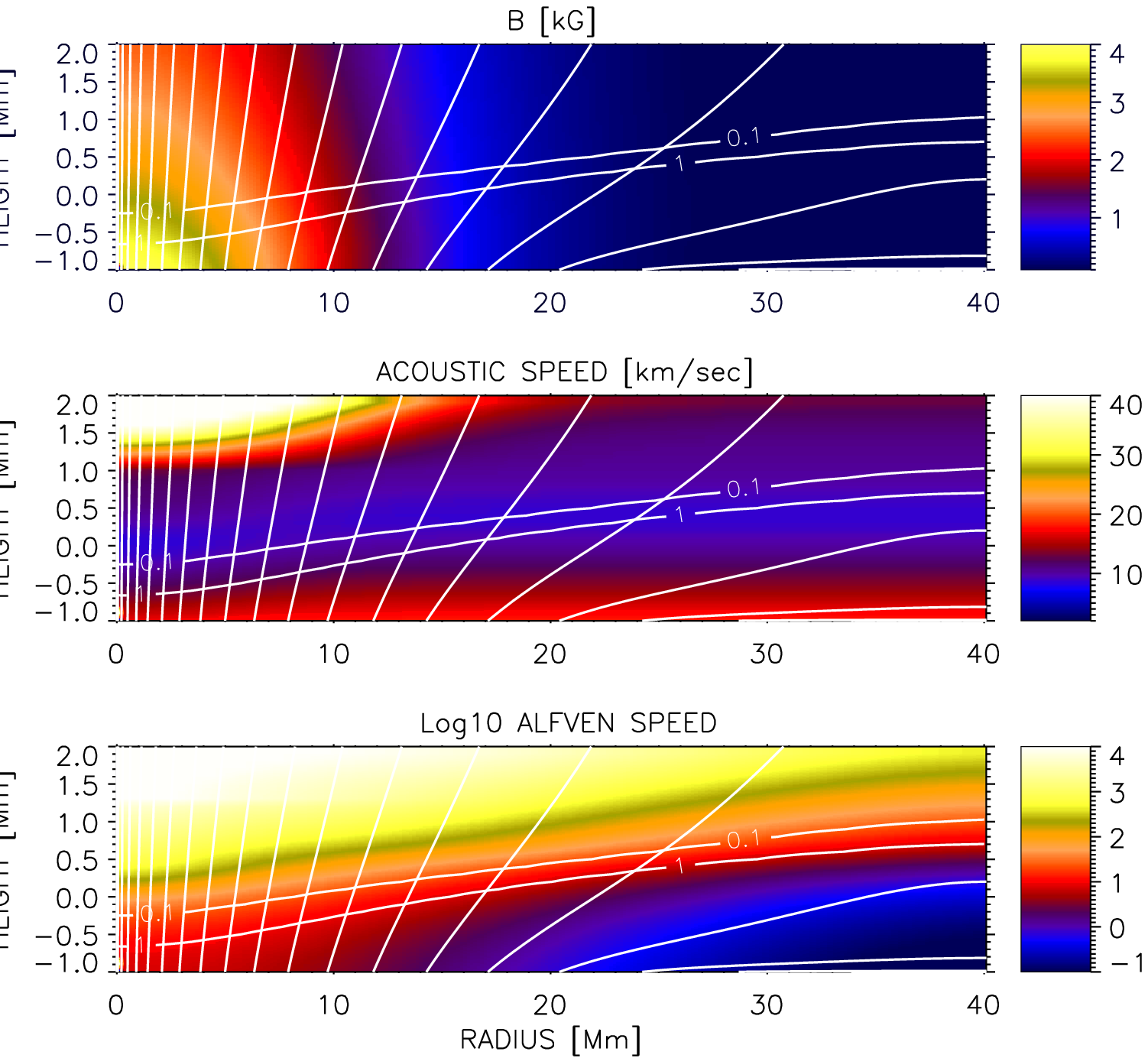}
\caption{Topology of the Pizzo solution with $B_0^P=4$ kG and
$r_e=9.4$ Mm. Top: magnetic field strength; middle: acoustic
speed; bottom: log of the Alfv\'en speed. White lines are magnetic
field lines. White lines with labels are the contours of
$c_S^2/v_A^2$. Note that, for better visualization, the vertical
axis has been expanded.} \label{fig:pizzo}
\end{figure}

The model sunspot for the parameters given above is shown in
Fig.~\ref{fig:pizzo}. The magnetic field on the axis drops from 4
kG at $z=-1$ Mm to about 2 kG at $=2$ Mm, which is a rather large
value at this height. Due to such a large field strength, the
Alfv\'en speed exceeds $10^4$ \kms\ in the upper layers. The image
of the sound speed shows the presence of the Wilson depression
around $z=0$, \ie\ the temperature at the sunspot axis is smaller
than in the outside atmosphere at a given height. Note that at
higher layers, the effect is the opposite and the temperature
inside the sunspot is larger. {This effect is due to initial
distribution in the model atmospheres taken as boundary
conditions. The field lines are more inclined comparing to the Low
solution in Fig.~\ref{fig:low}.

\subsection{Step 4: Concatenating the solutions}

\begin{figure}
\centering
\includegraphics[width=9cm]{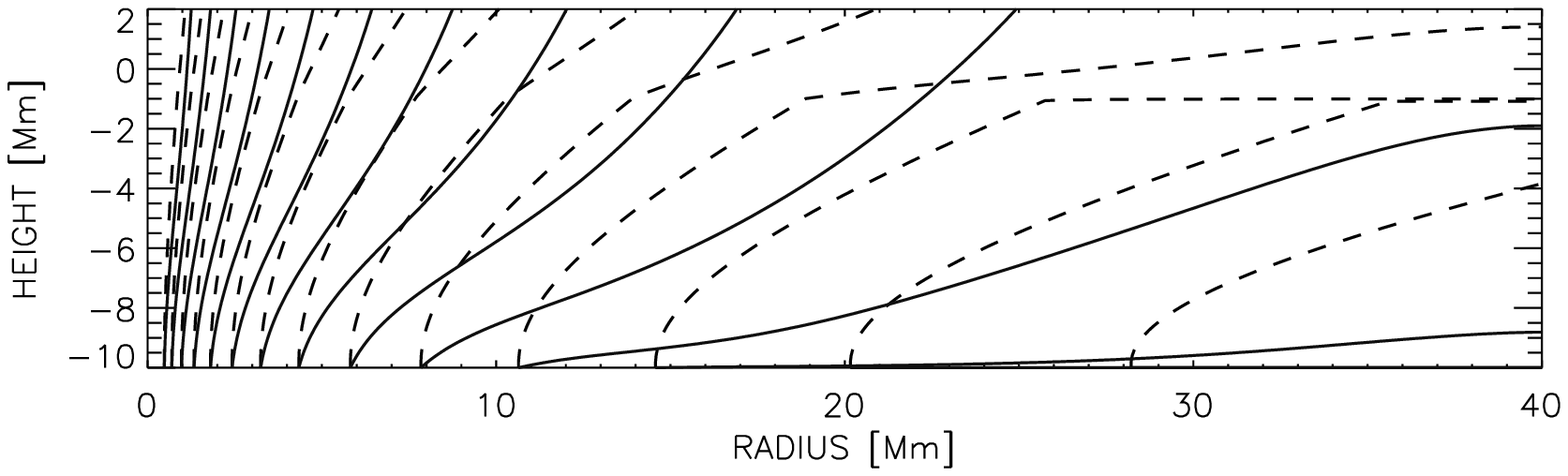}
\caption{Topology of the magnetic field lines before iterations
(dashed) and after a new equilibrium is reached (solid). }
\label{fig:cat}
\end{figure}
\begin{figure}
\centering
\includegraphics[width=9cm]{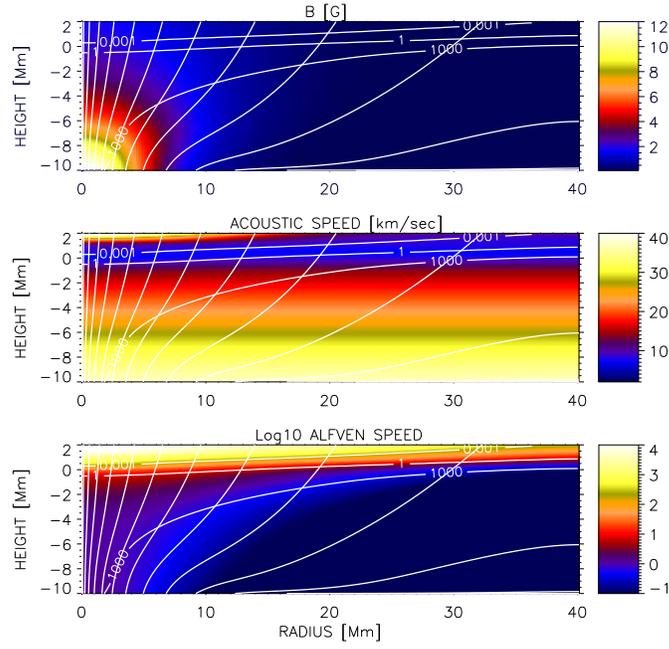}
\caption{Topology of the complete solution with $B_0^L=25000$,
$a=2h$, $h=3$ Mm and $\eta=1.3$ ($B_0^P=4$ kG and $r_e=9.4$ Mm).
Top: magnetic field strength; middle: acoustic speed; bottom: log
of the Alfv\'en speed. White lines are magnetic field lines. White
lines with labels are the contours of $c_S^2/v_A^2$. }
\label{fig:complete}
\end{figure}
\begin{figure}
\centering
\includegraphics[width=9cm]{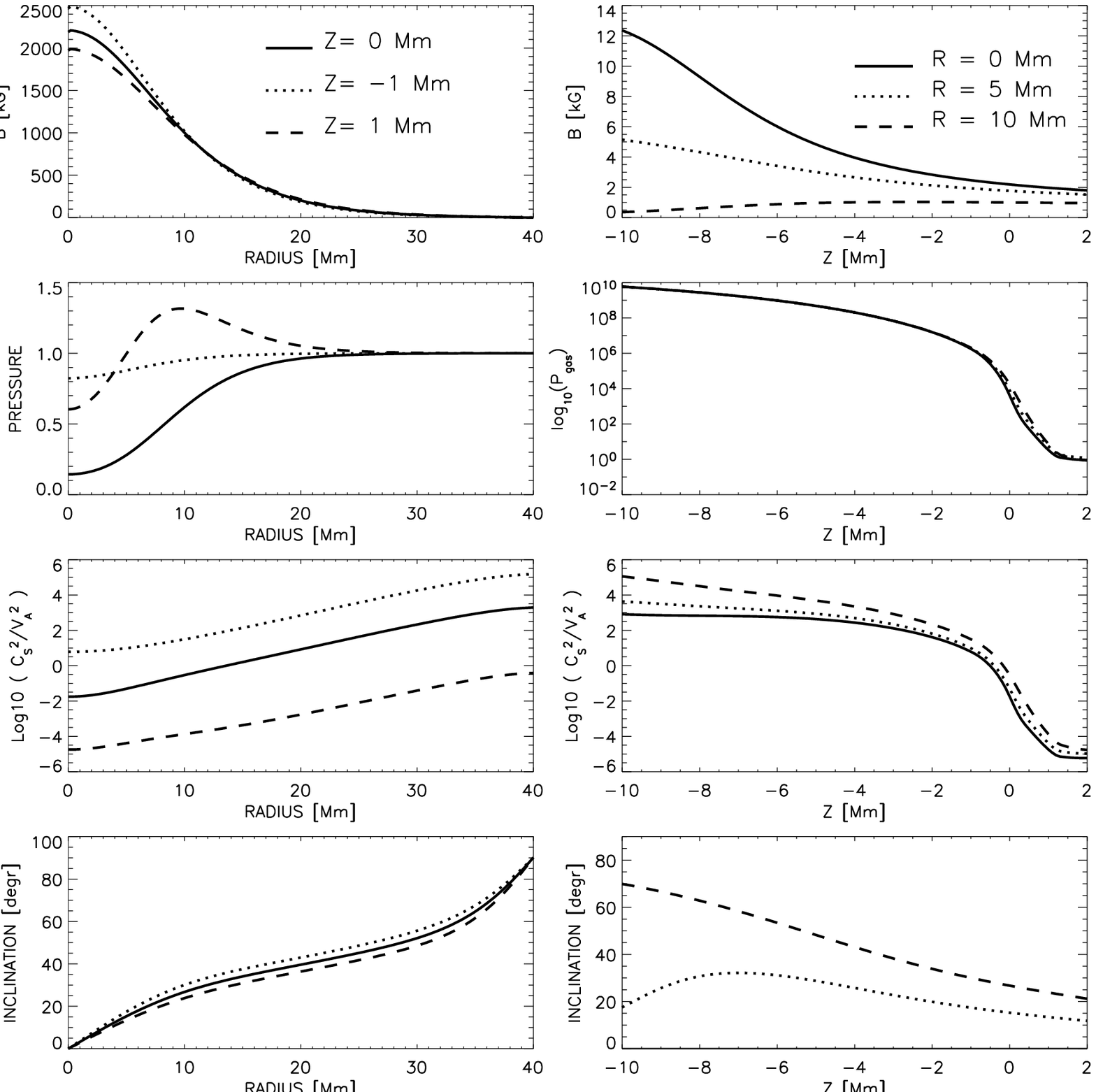}
\caption{Distribution with radial distance (left panels) and with
depth (right panels) of the magnetic field strength, pressure,
ratio $c_S^2/v_A^2$ and the magnetic field inclination for the
sunspot with $B_0^L=25000$, $a=2h$, $h=3$ Mm and $\eta=1.3$
(corresponding to $B_0^P=4$ kG and $r_e=9.4$ Mm at $z= -1$ Mm).
The radial pressure distributions are normalized to their values
at the right boundary. } \label{fig:radius}
\end{figure}

Both solutions obtained in Step 1 and Step 3 are in MHS
equilibrium. In order to construct the complete model from deep to
high layers one has to put one model on top of the other.
However, despite $B_z$ at the bottom boundary of the Pizzo model
is calculated in agreement with $B_z$ at the top boundary of the
Low model, there is a discontinuity in the horizontal component of
the magnetic field.
This discontinuity can be appreciated in Fig.~\ref{fig:cat}, where
the field lines with a dashed style correspond to the two models
concatenated as they are.
The reason of this discontinuity is two-fold. On the one hand, the
physics of the solution changes abruptly from one model to
another, thus changing the gradients of the magnetic field, gas
pressure, \etc \ On the other hand, the boundary condition for the
field line constant $u$ is not the same in the both models. In the
case of Low model, there is no need to put a condition on $u$,
neither there is a possibility. The inclination of the magnetic
field lines at the right boundary is a consequence of the
parameters of the model and should not be necessarily horizontal.
Contrarily, in the case of the Pizzo model, we impose horizontal
magnetic field at the right boundary.
The dependence of $B_z$ on $r$ given by Eq.~\ref{eq:bzpizzo}
defines the vertical magnetic field strength but does not put
constraints on the horizontal field component.

Thus, in order to obtain a smooth solution everywhere in the
domain, we repeat the Step 3 calculations for the complete model
sunspot. We take the pressure distributions at the axis and in the
field-free outside atmosphere from  the joint model at all
heights. The boundary conditions for $u$ are the same as in the
Pizzo model. The distribution of $u$ at the bottom boundary is
taken from the Low model. Then we repeat the solution of
Eq.~\ref{eq:pizzo}. The resulted topology of the magnetic field
lines is plotted in Fig.~\ref{fig:cat} in solid line style. The
field lines in the final solution are more horizontal in low
layers, while they are more vertical in the upper layers.

Fig.~\ref{fig:complete} gives the distribution of some parameters
in the complete model sunspot, at all layers. We can see that the
last iteration has re-distributed all the parameters compared to
the individual Low and Pizzo parts of the solution. In particular,
the magnetic field gradient at the axis is now more steep and the
field strength at high layers becomes lower. The field is  in
general more inclined, being horizontal at the right hand side
domain boundary, consistent with our imposed boundary condition
there. The gas pressure is modified accordingly to maintain the
new force balance.

Fig.~\ref{fig:radius} gives a more detailed view on the model spot
solution. It shows the distribution with radius (left panels) and
with depth (right panels) of some parameters of the sunspot
atmosphere. The field strength decreases rapidly with height at an
average rate of about 1 G/km at the axis. The magnitude of the
gradient decreases with height and with distance to the axis.
These gradients are in agreement with photospheric
spectropolarimetric observations \citep{Solanki2003}. The
magnitude of the pressure deficit inside the model sunspot
decreases with depth almost disappearing at about -2 Mm depth, in
accordance with our assumption of self-similarity of the MHS
solution at larger depths.

As can be seen from the radial pressure distribution, there is a
pressure excess observed at larger heights in the chromosphere at
some distance from the axis. This pressure excess would produce a
bright ring in the emergent intensity from the model sunspot and
it is present in the original model by \citet{Pizzo1986}. As shown
in the latter work, the bright ring can be removed by an improved
initial guess of the $P(u,z)$ distribution. It is unimportant for
the purpose of the present work since we only need an approximate
agreement between the average properties of the MHS solution and
the observed properties of sunspots. The magnetic field lines of
the model sunspot are inclined less than 30 degrees within the
first 10 Mm from the axis, which can be considered as the umbra.
Due to the boundary condition, the inclination changes gradually
becoming 90 degrees at the edge of the model, where the magnetic
field is already very weak.
The ratio between the sound speed and the Alfv\'en speed squared
(which gives the measure of the gas to magnetic pressure) changes
orders of magnitude from $10^6$ at $z=-10$ Mm to $10^{-6}$ at
$z=$2 Mm. Note that despite this, there is no problem with the
convergence of the solution.
%

In the next section, we give more examples of MHS solutions
comparing models obtained with various sets of parameters. In the
examples below we discuss the models calculated in a complete
domain from $z=-10$ to $z=2$ Mm.

\begin{figure}
\centering
\includegraphics[width=9cm]{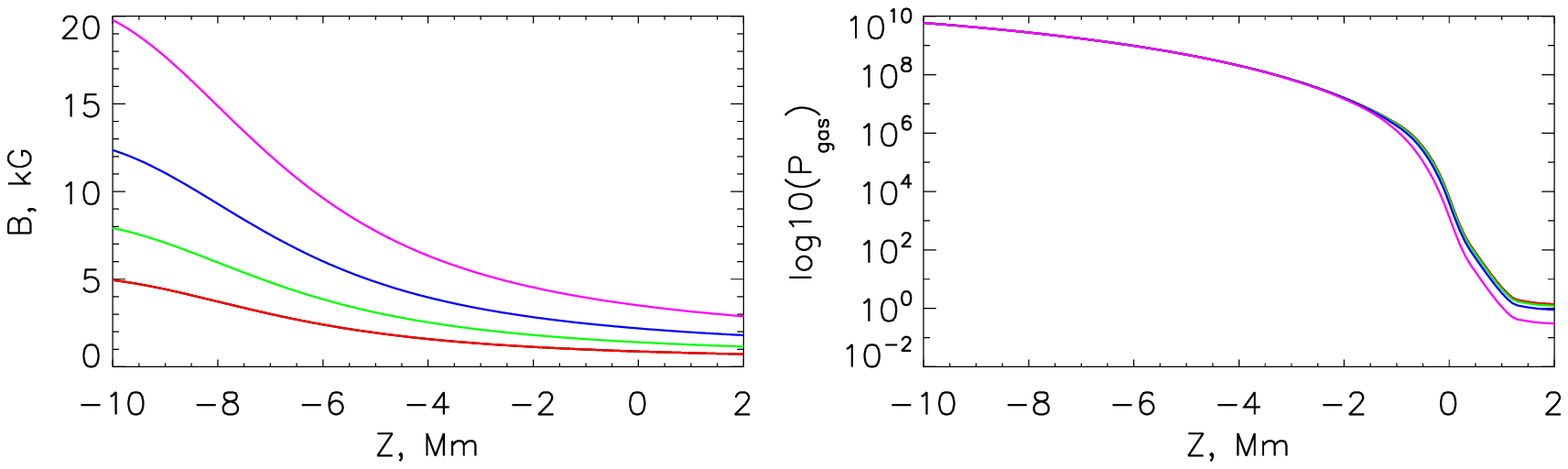}
\includegraphics[width=9cm]{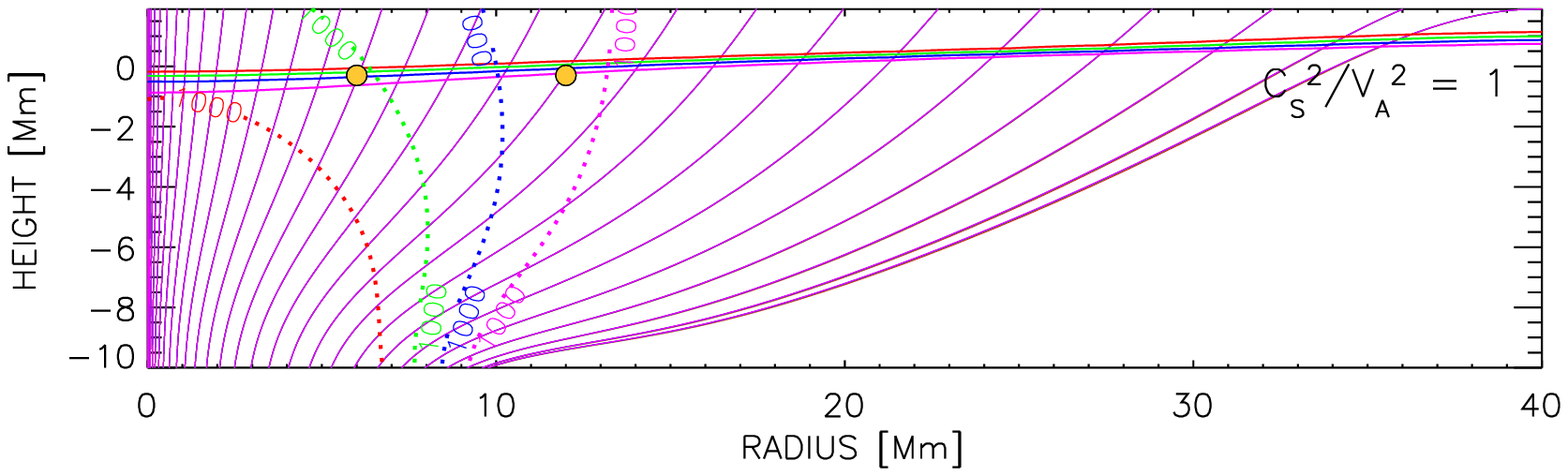}
\caption{Top panels: height dependence of the magnetic field and
pressure at the axis for the models with $a=2h$, $h=3$ Mm,
$\eta=1.3$ and $B_0^L$= 10000 (red line), 16000 (green  line),
25000 (blue line) and 40000 (magenta line) G. Bottom panel:
topology of the magnetic field lines for the same solutions (same
color coding). Contours of the magnetic field strength of $B=1000$
G are shown by dotted lines for each case. Horizontal solid lines
mark the levels of $c_S=v_A$. } \label{fig:height}
\end{figure}

\section{Examples}

{\bf Dependence on magnetic field strength.} Fig.~\ref{fig:height}
shows the magnetic field topology of models with different values
of the magnetic field strength (parameter $B_0^L$), all the other
parameters being exactly the same as previously.
As can be seen by comparing the different curves on the figures,
the resulting gas pressure stratifications only differ in the
highest layers by the amount of the pressure deficit.
The magnetic field topology is indistinguishable in all the cases.
This property originates from two effects. On the one side, the
self-similar solution in the bottom part of the domain scales with
magnetic field strength, i.e., the field line topology does not
depend on $B_0^L$.
On the other side, the Pizzo solution in the upper part is close
to potential, imposed by the solution at the bottom part. The
potential solution also scales with the magnetic field strength
and is independent of the thermodynamic properties.
Both effects lead to the independence of $B_0^L$  of the magnetic
topology of the final solution in the complete domain.
This is a useful property from the point of view of
helioseismology simulations. Using a set of models with different
magnetic field strength, but otherwise the same, the effects of
the magnetic field strength on waves can be checked independently
of the effects of the magnetic field inclination.

It should be noted that the above property originates only from
the particular choice of the parameters $a$ and $\eta$. This
choice produces $r_e$ large enough, so that the final solution in
the upper part of the domain approaches to potential and becomes
almost independent on the pressure distribution $P(u,z)$
\citep{Pizzo1986}.

The models presented in this Section are available electronically
in FITS format from the Astrophysical Journal web site.

\begin{figure}
\centering
\includegraphics[width=9cm]{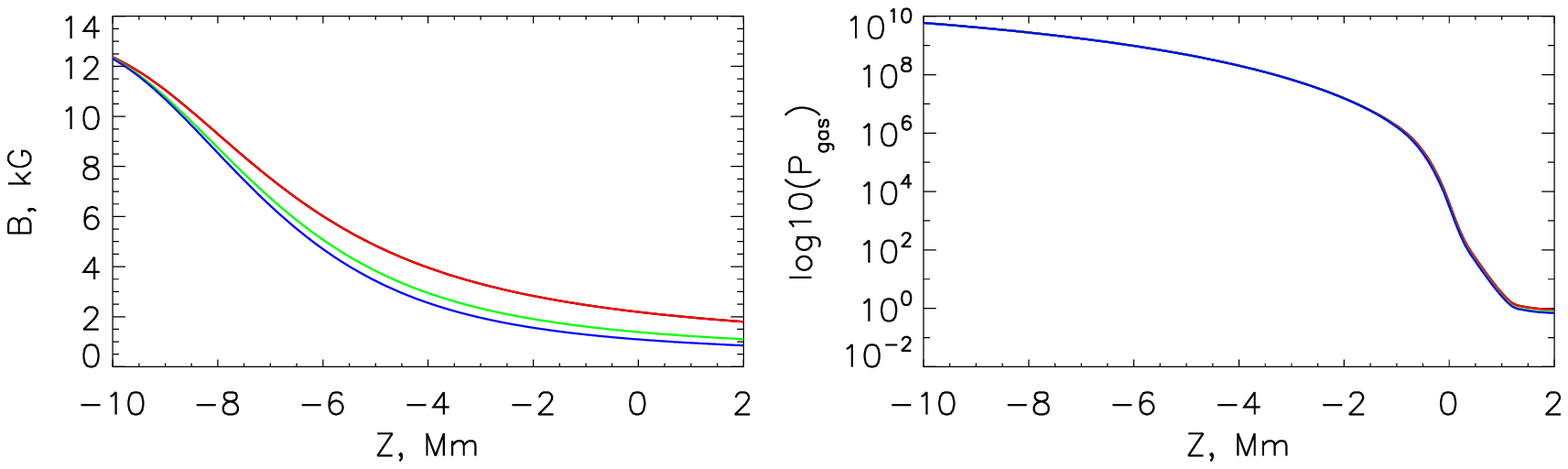}
\includegraphics[width=9cm]{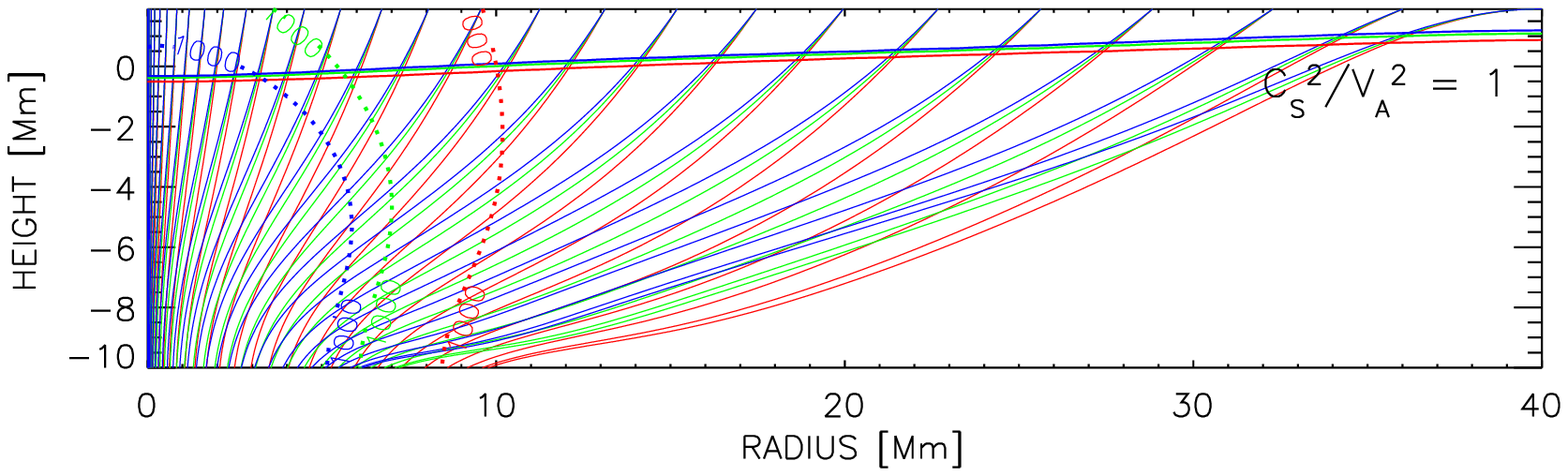}
\caption{Top panels: height dependence of the magnetic field and
pressure at the axis for the models with $a=2h$, $h=3$ Mm,
$B_0^L$= 25000 and $\eta=$ 1.3 (red line), 2.5 (green line) and
3.5 (blue line). Bottom panel: topology of the magnetic field
lines for the same solutions (same color coding). Contours of the
magnetic field strength of $B=1000$ G are shown by dotted lines
for each case. Horizontal solid lines mark the levels of
$c_S=v_A$.} \label{fig:height_eta}
\end{figure}

{\bf Dependence on $a$ and $\eta$.} Fig.~\ref{fig:height_eta}
shows the magnetic field topology of the models calculated with
different values of the parameter $\eta$ (Eqs.~\ref{eq:brlow} and
\ref{eq:bzlow}), the rest of the parameters being the same.
Note that, according to the above equations, the inclination of
the magnetic field is independent of $\eta$ in the Low model at
the bottom part of the domain. However, the initial radius of the
structure in the Pizzo part of the solution $r_e$
(Eq.~\ref{eq:ltp2}) depends on $\eta$, thus changing the
inclination of the magnetic field lines in the upper part of the
atmosphere. The final iteration performed in Step 4 takes that
into account, making the solution in the complete domain dependent
on $\eta$.

The change of $\eta$ produces two effects. By increasing $\eta$,
we decrease the magnetic field strength by a smaller amount than
by varying $B_0^L$, as in the previous example. At the same time
increasing $\eta$ produces an increase of the inclination of the
magnetic field lines in the solution in the complete domain. The
difference in the inclination is more pronounced in the deep
layers of the model.
The magnetic field topology of the solutions is different. The
gradient of the magnetic field at the axis is slightly larger for
larger values of $\eta$ in the sub-photospheric part of the model.
Again, this difference is produced after the final iteration in
Step 4 since the $B_z(r=0)$ given by Eq.~\ref{eq:bzlow} of the Low
part of the solution is independent of $\eta$. The pressure
distribution at the axis and the amount of the pressure deficit
are not very different between the given models. The models
presented in this Section are available electronically in FITS
format from the Astrophysical Journal web site.

Varying the parameter $a$ produces similar effects. The difference
is that, by varying $a$, we change mostly the curvature of the
magnetic field lines and the radius of the structure, not
affecting much the magnetic field strength.

\begin{figure}
\centering
\includegraphics[width=9cm]{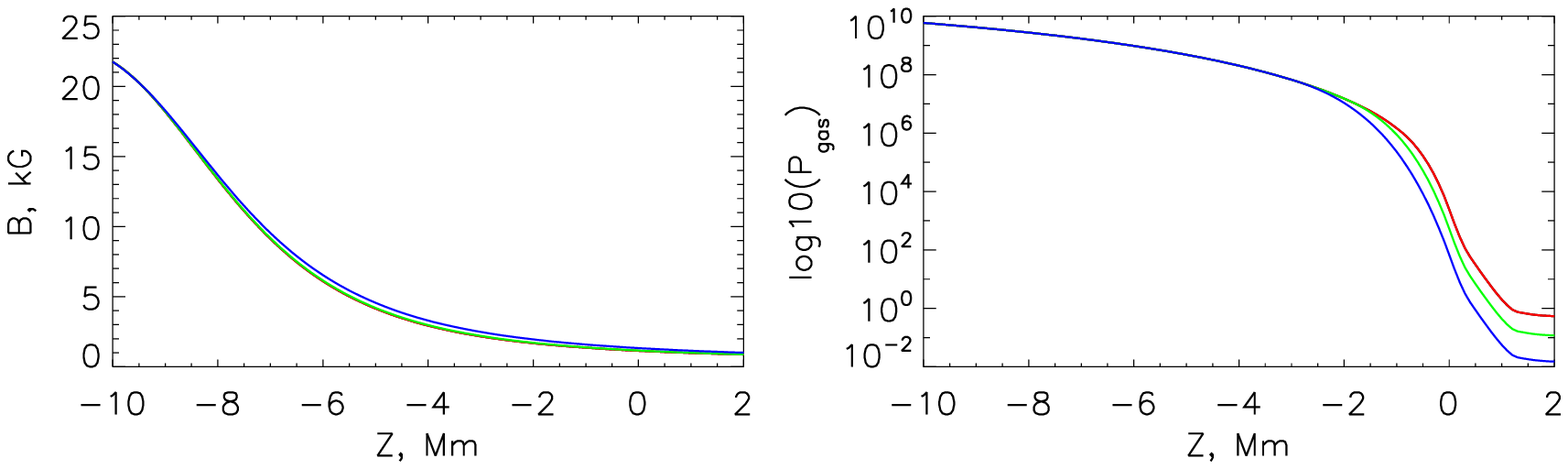}
\includegraphics[width=9cm]{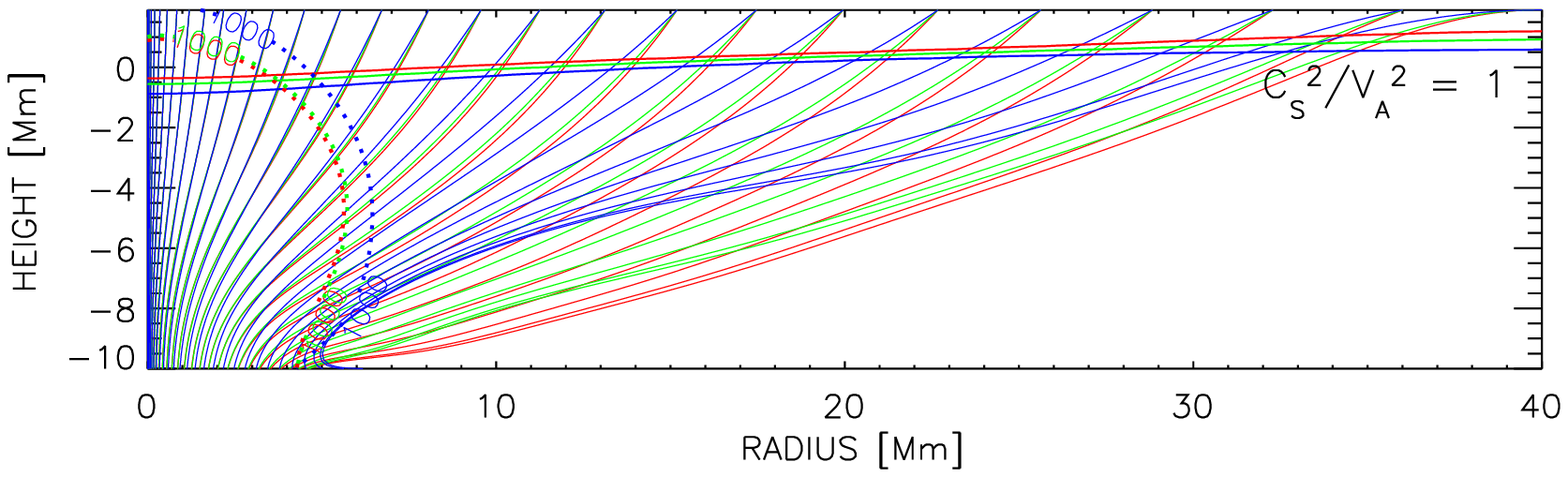}
\caption{Top panels: height dependence of the magnetic field and
pressure at the axis for the models with $a=1.5h$, $h=3$ Mm,
$B_0^L$= 25000, $\eta=$ 3.5 and $z_0$=-1 Mm (red line), -2 Mm
(green line) and -3 Mm (blue line). Bottom panel: topology of the
magnetic field lines for the same models (same color coding).
Contours of the magnetic field strength of $B=1000$ G are shown by
dotted lines for each case. Horizontal solid lines mark the levels
of $c_S=v_A$. } \label{fig:height_zdeep}
\end{figure}

{\sc \bf Dependence on $z_0$.} Another parameter introduced in our
modeling is the height where both solutions merge, $z_0$. Fig.
\ref{fig:height_zdeep} shows the topology of the magnetic field
lines and the pressure and density distributions with height at
the axis, for three models with different values of $z_0$. In this
example we take different values of $a$ and $\eta$ compared to
above examples (see figure caption) in order to produce a
structure with a smaller radius. This way, we show that the
procedure is robust and can produce magnetic structures with very
different properties.

The magnetic field strength at the axis is almost independent of
the choice of $z_0$. The amount of the pressure deficit at the
near-surface layers increases with decreasing $z_0$ from $-1$ to
$-3$ Mm, extending to larger depths.
Note, however, that we can not shift the level of $z_0$ much
deeper than $-3$ Mm due to a poor convergence of the solution.
Despite the magnetic field strength is nearly the same, the
position of the $\beta=1$ level is different in all the solutions
due to the different amount of the pressure deficit.
The inclination of the magnetic field lines is similar in the
central part of the model sunspots in the three solutions. At the
periphery, especially at larger depths, the field lines get more
inclined with decreasing $z_0$. Thus, the field is more
concentrated to the central part and the effective radius of the
structure is smaller.
The models presented in this Section are available electronically
in FITS format from the Astrophysical Journal web site.


\section{Conclusions}

In this study, we propose a method to construct a magnetostatic
structure with properties and size of a typical sunspot, from the
deep interior to the solar surface.
Previously published methods to construct such a model fail due to
a poor knowledge of the thermodynamic and magnetic parameters of
sunspots in sub-photospheric layers.
We make use of self-similar models in deep layers and show that
such models can naturally merge with models where the pressure
distribution is prescribed on the axis, as well as the field-free
atmosphere, allowing for a more realistic description of the
atmospheric layers of sunspots.
The procedure shows a rather good convergence and stability. By
changing the parameters of the solution, a set of models can be
produced with desired properties.
We suggest that these models may be used, among others, in
artificial helioseismology data simulations. Given a set of
models, a parametric study can be done, investigating the
influence of the topology and strength of the magnetic field of
sunspots on the parameters inferred by the local helioseismology
measurements in solar active regions.

\begin{acknowledgements}
      This research has been funded by the Spanish
Ministerio de Educaci{\'o}n y Ciencia through projects
AYA2007-63881 and AYA2007-66502.
\end{acknowledgements}


\end{document}